# A TWO-STEP TIME OF ARRIVAL ESTIMATION ALGORITHM FOR IMPULSE RADIO ULTRA WIDEBAND SYSTEMS[1]

*Sinan Gezici\*, Zafer Sahinoglu†, Andreas F. Molisch†,2 Hisashi Kobayashi\*, and H. Vincent Poor\**

\* Department of Electrical Engineering
Princeton University, Princeton, NJ 08544
phone:+(1)-609-258-6868
fax:+(1)-609-258-2158
{sgezici,hisashi,poor}@princeton.edu

† Mitsubishi Electric Research Laboratories
201 Broadway, Cambridge, MA 02139
phone:+(1)-617-621-7500
fax:+(1)-617-621-7550
{zafer,molisch}@merl.com

## ABSTRACT

High time resolution of ultra wideband (UWB) signals facilitates very precise positioning capabilities based on time-of-arrival (TOA) measurements. Although the theoretical lower bound for TOA estimation can be achieved by the maximum likelihood principle, it is impractical due to the need for extremely high-rate sampling and the presence of large number of multipath components. On the other hand, the conventional correlation-based algorithm, which serially searches possible signal delays, takes a very long time to estimate the TOA of a received UWB signal. Moreover, the first signal path does not always have the strongest correlation output. Therefore, first path detection algorithms need to be considered. In this paper, a data-aided two-step TOA estimation algorithm is proposed. In order to speed up the estimation process, the first step estimates the rough TOA of the received signal based on received signal energy. Then, in the second step, the arrival time of the first signal path is estimated by considering a hypothesis testing approach. The proposed scheme uses low-rate correlation outputs, and is able to perform accurate TOA estimation in reasonable time intervals. The simulation results are presented to analyze the performance of the estimator.

## 1. INTRODUCTION

Since the US Federal Communications Commission (FCC) approved the limited use of UWB technology [1], communications systems that employ UWB signals have drawn considerable attention. A UWB signal is defined to be one that possesses an absolute bandwidth larger than 500MHz or a relative bandwidth larger than 20%. It can coexist with incumbent systems in the same frequency range due to its large spreading factor and low power spectral density. UWB technology holds great promise for a variety of applications such as short-range high-speed data transmission and precise location estimation.

Commonly, impulse radio (IR) systems, which transmit very short pulses with a low duty cycle, are employed to implement UWB systems ([2]-[6]). In an IR system, a train of pulses is sent and information is usually conveyed by the position or the polarity of the pulses, which correspond to Pulse Position Modulation (PPM) and Binary Phase Shift Keying (BPSK), respectively. In order to prevent catastrophic collisions among different users and thus provide robustness against multiple-access interference, each information symbol is represented by a sequence of pulses; the positions of the pulses within that sequence are determined by a pseudo-random time-hopping (TH) sequence specific to each user [2].

The high time resolution of UWB signals facilitates very precise TOA estimation, as suggested by the Cramer-Rao lower bound. However, in practical systems, the challenge is to perform this estimation in a reasonable time interval using low-rate correlation outputs.

The conventional correlation-based TOA estimation algorithms are both suboptimal and require exhaustive search among thousands of bins, which results in very slow TOA estimation [7]. In order to speed up the process, different search strategies, such as random search or bit reversal search, are proposed in [8]. In [9], a generalized maximum likelihood (GML) estimation principle is employed to obtain iterative solutions after some simplifications. However, this approach requires very high rate sampling, which is not practical in many applications. An alternative to the GML-based approach is a low complexity non-data-aided timing offset estimation technique based on *symbol-rate* samples based on the novel idea of "dirty templates" [10]-[13]. The main disadvantage of the timing with dirty templates (TDT) algorithm is that its TOA estimate will have an ambiguity equal to the extent of the noise-only region between consecutive symbols.

In this paper, we propose a two-step TOA estimation algorithm. In order to speed up the estimation process, the first step estimates the rough TOA of the received signal based on received signal energy. Then, in the second step, the arrival time of the first signal path is estimated from low-rate correlation outputs by considering a hypothesis testing approach.

The remainder of the paper is organized as follows. Section II describes the transmitted and received signal models in a frequency-selective environment. The two-step TOA estimation algorithm is considered in Section III, which is followed by simulation results in Section IV, and concluding remarks are made in the last section.

---
[1]This research was supported by Mitsubishi Electric Research Laboratories, Cambridge, MA, in part by the National Science Foundation under grant ANI-03-38807, in part by the Army Research Laboratory under contract DAAD 19-01-2-0011, and in part by the New Jersey Center for Wireless Telecommunications.

[2]Also at the Department of Electroscience, Lund University, Sweden.

## 2. SIGNAL MODEL

We consider a BPSK TH-IR transmitted signal represented by:

$$s_{\text{tx}}(t) = \sqrt{E} \sum_{j=-\infty}^{\infty} a_j \, b_{\lfloor j/N_f \rfloor} w_{\text{tx}}(t - jT_f - c_j T_c), \quad (1)$$

where $w_{\text{tx}}(t)$ is the transmitted UWB pulse with duration $T_c$, $E$ is the transmitted pulse energy, $T_f$ is the "frame" time, and $b_{\lfloor j/N_f \rfloor} \in \{+1, -1\}$ is the binary information symbol. In order to smooth the power spectrum of the transmitted signal and allow the channel to be shared by many users without causing catastrophic collisions, a time-hopping (TH) sequence $c_j \in \{0, 1, ..., N_c - 1\}$ is assigned to each user, where $N_c$ is the number of chips per frame interval; that is, $N_c = T_f/T_c$. Additionally, random polarity codes, $a_j$'s, can be employed, which are binary random variables taking on the values $\pm 1$ with equal probability, and are known to the receiver. Use of random polarity codes helps reduce the spectral lines in the power spectral density of the transmitted signal [14] and mitigate the effects of MAI [15].

Consider the following channel model

$$h(t) = \sum_{l=1}^{L} \alpha_l \delta(t - (l-1)T_c - \tau_{\text{TOA}}), \quad (2)$$

where $\alpha_l$ is the channel coefficient for the $l$th path, $L$ is the number of multipath components, and $\tau_{\text{TOA}}$ is the TOA of the incoming signal.

From (1) and (2), we can express the received signal as

$$r(t) = \sum_{l=1}^{L} \sqrt{E} \, \alpha_l s_{\text{rx}}(t - (l-1)T_c - \tau_{\text{TOA}}) + n(t), \quad (3)$$

where $s_{\text{rx}}(t)$ is given by

$$s_{\text{rx}}(t) = \sum_{j=-\infty}^{\infty} a_j \, b_{\lfloor j/N_f \rfloor} w_{\text{rx}}(t - jT_f - c_j T_c), \quad (4)$$

with $w_{\text{rx}}(t)$ denoting the received UWB pulse with unit energy.

Assuming a data aided TOA estimation scheme, we consider a training sequence of $b_j = 1 \, \forall j$. Then, (4) can be expressed as

$$s_{\text{rx}}(t) = \sum_{j=-\infty}^{\infty} a_j w_{\text{rx}}(t - jT_f - c_j T_c). \quad (5)$$

We assume, for simplicity, that the signal always arrives in one frame duration ($\tau_{\text{TOA}} < T_f$), and there is no interframe interference (IFI); that is, $T_f \geq (L + c_{\max})T_c$ (equivalently, $N_c \geq L + c_{\max}$), where $c_{\max}$ is the maximum value of the TH sequence. Note that the assumption of $\tau_{\text{TOA}} < T_f$ does not restrict the validity of the algorithm. In fact, it is enough to have $\tau_{\text{TOA}} < T_s$ for the algorithm to work when the frame is large enough and predetermined TH codes are employed. Moreover, even if $\tau_{\text{TOA}} \geq T_s$, an initial energy detection can be used to determine the arrival time within a symbol uncertainty before running the proposed algorithm.

## 3. TWO-STEP TOA ESTIMATION ALGORITHM

One of the most challenging issues in UWB TOA estimation is to obtain a reliable estimate in a reasonable time interval under the constraint of sampling rate. In order to have a low power and low complexity receiver, one should assume low sampling rates at the output of the correlators. However, when low rate samples are employed, the TOA estimation can take a very long time. Therefore, we propose a novel two-step TOA estimation algorithm that can perform TOA estimation from low rate samples (typically on the order of hundreds times slower sampling rate than chip-rate sampling) in a reasonable time interval. In order to speed up the estimation process, the first step estimates the coarse TOA of the received signal based on received signal energy. Then, in the second step, the arrival time of the first signal path is estimated by considering a hypothesis testing approach.

Express the TOA $\tau_{\text{TOA}}$ in (3) as[3]

$$\tau_{\text{TOA}} = kT_c = k_b T_b + k_c T_c, \quad (6)$$

where $k \in [0, N_c - 1]$ is the TOA in terms of the chip interval $T_c$, $T_b$ is the block interval consisting of $B$ chips ($T_b = BT_c$), and $k_b \in [0, N_c/B - 1]$ and $k_c \in [0, B - 1]$ are the integers that determine, respectively, in which block and chip the first signal path arrives.

The two-step TOA algorithm first estimates the block in which the first signal path exists; then it estimates the chip position in which the first path resides. In other words, it can be summarized as:

- Estimate $k_b$ from received signal strength (RSS) measurements.
- Estimate $k_c$ (equivalently, $k$) from low rate correlation outputs using a hypothesis testing approach.

### 3.1. First Step: Coarse TOA Estimation from RSS Measurements

In the first step, the aim is to detect the *coarse* arrival time of the signal in the frame interval. Assume, without loss of generality, that the frame time $T_f$ is an integer multiple of $T_b$, the block size of the algorithm; that is, $T_f = N_b T_b$.

In order to have reliable decision variables in this step, energy is combined from $N_1$ different frames of the incoming signal for each block. Hence, the decision variables are expressed as

$$Y_i = \sum_{j=0}^{N_1-1} Y_{i,j}, \quad (7)$$

for $i = 0, \ldots, N_b - 1$, where

$$Y_{i,j} = \int_{jT_f + iT_b + c_j T_c}^{jT_f + (i+1)T_b + c_j T_c} |r(t)|^2 dt. \quad (8)$$

Then, $k_b$ in (6) is estimated as

$$\hat{k}_b = \arg\max_{0 \leq i \leq N_b - 1} Y_i. \quad (9)$$

---

[3]For simplicity, the TOA is assumed to be an integer multiple of the chip duration $T_c$. In a practical scenario, sub-chip resolution can be obtained by employing a delay-lock-loop (DLL) after the TOA estimation with chip-level uncertainty [16].

In other words, the block with the largest signal energy is selected.

The parameters of this step that should be selected appropriately are the block size $T_b$ ($N_b$) and the number of frames $N_1$, from which energy is collected.

## 3.2. Second Step: Fine TOA Estimation from Low Rate Correlation Outputs

After determining the coarse arrival time from the first step, the second step tries to estimate $k_c$ in (6). Ideally, $k_c \in [0, B-1]$ needs to be searched for TOA estimation, which corresponds to searching $k \in [\hat{k}_b B, (\hat{k}_b+1)B - 1]$, with $\hat{k}_b$ obtained from (9). However, in some cases, the first signal path can reside in one of the blocks prior to the strongest one due to multipath effects. Therefore, instead of searching a single block, $k \in [\hat{k}_b B - M_1, (\hat{k}_b+1)B - 1]$, with $M_1 \geq 0$, can be searched for the TOA in order to increase the probability of detection of the first path. In other words, in addition to the block with the largest signal energy, we can perform an additional backwards search over $M_1$ chips. For notational simplicity, let $\mathcal{U} = \{n_s, n_s+1, \ldots, n_e\}$ denote the uncertainty region, where $n_s = \hat{k}_b B - M_1$ and $n_e = (\hat{k}_b+1)B - 1$ are the start and end points.

In order to estimate the TOA with chip-level resolution, we consider correlations of the received signal with shifted versions of a template signal. For delay $iT_c$, we obtain the following correlation output

$$z_i = \int_{iT_c}^{iT_c + N_2 T_f} r(t) s_{\text{temp}}(t - iT_c) \, dt, \qquad (10)$$

where $N_2$ is the number of frames over which the correlation output is obtained, and $s_{\text{temp}}(t)$ is the template signal given by

$$s_{\text{temp}}(t) = \sum_{j=0}^{N_2-1} a_j w_{\text{rx}}(t - jT_f - c_j T_c). \qquad (11)$$

From the correlation outputs for different delays, the aim is to determine the chip in which the first signal path has arrived. By appropriate choice of the block interval $T_b$ and $M_1$, and considering a large number of multipath components in the received signal, which is typical for indoor UWB systems, we can assume that the block starts with a number of chips with noise-only components and the remaining ones with signal plus noise components, as shown in Figure 1. Assuming that the statistics of the signal paths do not change significantly in the uncertainty region $\mathcal{U}$, we can express the different hypotheses approximately as follows:

$$\begin{aligned}
\mathcal{H}_0 &: \quad z_i = \eta_i, & i &= n_s, \ldots, n_f, \\
\mathcal{H}_k &: \quad z_i = \eta_i, & i &= n_s, \ldots, k-1, \\
& \quad z_i = N_2 \sqrt{E} \, \alpha_{i-k+1} + \eta_i, & i &= k, \ldots, n_f, \quad (12)
\end{aligned}$$

for $k \in \mathcal{U}$, where $\eta_i$'s denote the i.i.d. output noise distributed as $\mathcal{N}(0, \sigma_n^2)$ with $\sigma_n^2 = N_2 \mathcal{N}_0 / 2$, $\alpha_1, \ldots, \alpha_{n_f - k + 1}$ are independent channel coefficients, assuming $n_f - n_s + 1 \leq L$, and $n_f = n_e + M_2$ with $M_2$ being the number of correlation outputs that are considered out of the uncertainty region in order to have reliable estimates of the unknown parameters of $\alpha$.

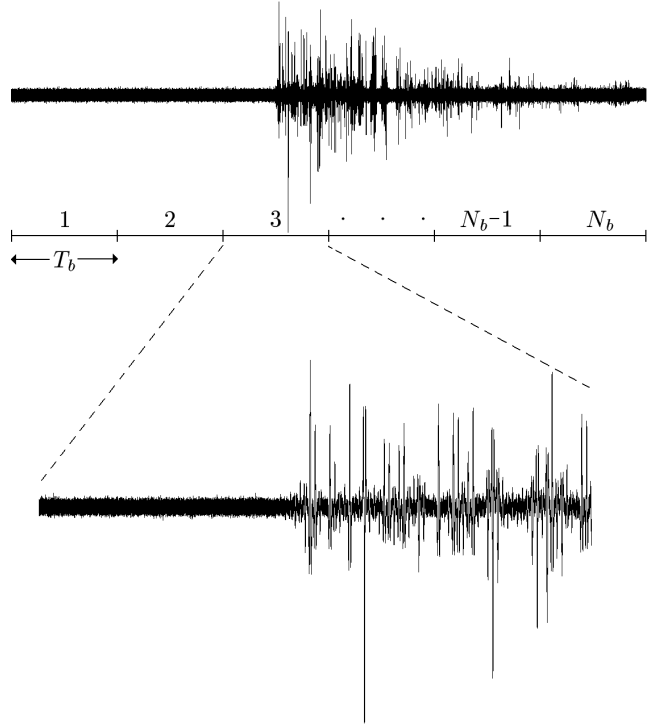

**Fig. 1**. Illustration of the two-step TOA estimation algorithm. The signal on the top is the received signal in one frame. The first step checks the signal energy in $N_b$ blocks and chooses the one with the highest energy (Although one frame is shown in the figure, energy from different frames can be collected for reliable decisions). Assuming that the third block has the highest energy, the second step focuses on this block (or an extension of that) to estimate the TOA. The zoomed version of the signal in the third block is shown on the bottom.

Due to very high resolution of UWB signals, it is appropriate to model the channel coefficients approximately as

$$\begin{aligned}
\alpha_1 &= d_1 |\alpha_1|, \\
\alpha_l &= \begin{cases} d_l |\alpha_l|, & p \\ 0, & 1-p \end{cases}, \quad l = 2, \ldots, n_f - n_s + 1,
\end{aligned} \qquad (13)$$

where $p$ is the probability that a channel tap arrives in a given chip, $d_l$ is the sign of $\alpha_l$, which is $\pm 1$ with equal probability, and $|\alpha_l|$ is the amplitude of $\alpha_l$, which is modelled as a Nakagami-$m$ distributed random variable with parameter $\Omega$; that is [17],

$$p(\alpha) = \frac{2}{\Gamma(m)} \left(\frac{m}{\Omega}\right)^m \alpha^{2m-1} e^{-\frac{m\alpha^2}{\Omega}}, \qquad (14)$$

for $\alpha \geq 0$, $m \geq 0.5$ and $\Omega \geq 0$, where $\Gamma$ is the Gamma function [18].

From the formulation in (12), it is observed that the TOA estimation problem can be considered as a "change detection" problem [19]. Let $\boldsymbol{\theta}$ denote the unknown parameters of the distribution of $\alpha$; that is, $\boldsymbol{\theta} = [p \ m \ \Omega]$. Then, the

log-likelihood ratio (LLR) is calculated as

$$S_k^{n_f}(\boldsymbol{\theta}) = \sum_{i=k}^{n_f} \log \frac{p_{\boldsymbol{\theta}}(z_i|\mathcal{H}_k)}{p(z_i|\mathcal{H}_0)}, \quad (15)$$

where $p_{\boldsymbol{\theta}}(z_i|\mathcal{H}_k)$ denotes the probability density function (p.d.f.) of the correlation output under hypothesis $\mathcal{H}_k$ and with unknown parameters given by $\boldsymbol{\theta}$, and $p(z_i|\mathcal{H}_0)$ denotes the p.d.f. of the correlation output under hypothesis $\mathcal{H}_0$.

Since $\boldsymbol{\theta}$ is unknown, its ML estimate can be obtained first for a given hypothesis $\mathcal{H}_k$ and then that estimate can be used in the LLR expression. In other words, the generalized LLR approach [19] can be taken, where the TOA estimate is expressed as

$$\hat{k} = \arg\max_{k \in \mathcal{U}} S_k^{n_f}(\hat{\boldsymbol{\theta}}_{ML}(k)), \quad (16)$$

where

$$\hat{\boldsymbol{\theta}}_{ML}(k) = \arg\sup_{\boldsymbol{\theta}} S_k^{n_f}(\boldsymbol{\theta}). \quad (17)$$

However, the ML estimate is usually very complex to calculate. Therefore, simpler estimators such as the method of moments (MM) estimator can be employed to obtain those parameters. The $n$th moment of a random variable $X$ having Nakagami-$m$ distribution with parameter $\Omega$ is given by

$$E\{X^n\} = \frac{\Gamma(m+n/2)}{\Gamma(m)} \left(\frac{\Omega}{m}\right)^{n/2}. \quad (18)$$

Then, from the correlator outputs $\{z_i\}_{i=k+1}^{n_f}$, the MM estimates for the unknown parameters can be obtained after some manipulation:

$$p_{MM} = \frac{\gamma_1 \gamma_2}{2\gamma_2^2 - \gamma_3}, \ m_{MM} = \frac{2\gamma_2^2 - \gamma_3}{\gamma_3 - \gamma_2^2}, \ \Omega_{MM} = \frac{2\gamma_2^2 - \gamma_3}{\gamma_2}, \quad (19)$$

where

$$\gamma_1 \triangleq \frac{1}{EN_2^2}(\mu_2 - \sigma_n^2),$$
$$\gamma_2 \triangleq \frac{1}{E^2 N_2^4} \left(\frac{\mu_4 - 3\sigma_n^4}{\gamma_1} - 6EN_2^2\sigma_n^2\right),$$
$$\gamma_3 \triangleq \frac{1}{E^3 N_2^6} \left(\frac{\mu_6 - 15\sigma_n^6}{\gamma_1} - 15E^2 N_2^4 \gamma_2 \sigma_n^2 - 45EN_2^2\sigma_n^4\right), \quad (20)$$

with $\mu_j$ denoting the $j$th sample moment given by

$$\mu_j = \frac{1}{n_f - k} \sum_{i=k+1}^{n_f} z_i^j. \quad (21)$$

Then, the index of the chip having the first signal path can be obtained as

$$\hat{k} = \arg\max_{k \in \mathcal{U}} S_k^{n_f}(\hat{\boldsymbol{\theta}}_{MM}(k)), \quad (22)$$

where $\boldsymbol{\theta}_{MM}(k) = [p_{MM} \ m_{MM} \ \Omega_{MM}]$ is the MM estimate for the unknown parameters[4].

---
[4]Note that the dependence of $p_{MM}$, $m_{MM}$ and $\Omega_{MM}$ on the change position $k$ is not shown explicitly for notational simplicity.

Let $p_1(z)$ and $p_2(z)$, respectively, denote the distributions of $\eta$ and $N_2\sqrt{E}\,d|\alpha| + \eta$. Then, the generalized LLR for the $k$th hypothesis is given by

$$S_k^{n_f}(\hat{\boldsymbol{\theta}}) = \log \frac{p_2(z_k)}{p_1(z_k)} + \sum_{i=k+1}^{n_f} \log \frac{p\,p_2(z_i) + (1-p)p_1(z_i)}{p_1(z_i)}, \quad (23)$$

where

$$p_1(z) = \frac{1}{\sqrt{2\pi}\sigma_n} e^{-\frac{z^2}{2\sigma_n^2}} \quad (24)$$

and

$$p_2(z) = \frac{\nu_1}{\sqrt{2\pi}\sigma_n} e^{-\frac{z^2}{2\sigma_n^2}} \Phi\left(m, \frac{1}{2}; \frac{z^2}{\nu_2}\right), \quad (25)$$

with

$$\nu_1 \triangleq \frac{2\sqrt{\pi}\,\Gamma(2m)}{\Gamma(m)\Gamma(m+0.5)} \left(4 + \frac{2EN_2^2\Omega}{m\sigma_n^2}\right)^{-m},$$
$$\nu_2 \triangleq 2\sigma_n^2\left(1 + 2m\frac{\sigma_n^2}{EN_2^2\Omega}\right), \quad (26)$$

and $\Phi$ denoting a confluent hypergeometric function given by [18]

$$\Phi(\beta_1, \beta_2; x) = 1 + \frac{\beta_1}{\beta_2}\frac{x}{1!} + \frac{\beta_1(\beta_1+1)}{\beta_2(\beta_2+1)}\frac{x^2}{2!}$$
$$+ \frac{\beta_1(\beta_1+1)(\beta_1+2)}{\beta_2(\beta_2+1)(\beta_2+2)}\frac{x^3}{3!} + \cdots \quad (27)$$

Note that the p.d.f. of $N_2\sqrt{E}\,d|\alpha| + \eta$, $p_2(z)$, is obtained from (14), (24) and the fact that $d$ is $\pm 1$ with equal probability.

After some manipulation, the TOA estimation rule can be expressed as

$$\hat{k} = \arg\max_{k \in \mathcal{U}} \Bigg\{ \log\left[\nu_1 \Phi\left(m, 0.5; \frac{z_k^2}{\nu_2}\right)\right]$$
$$+ \sum_{i=k+1}^{n_f} \log\left[p\,\nu_1\Phi\left(m, 0.5; \frac{z_i^2}{\nu_2}\right) + 1 - p\right] \Bigg\}. \quad (28)$$

### 3.3. Additional Tests

Note that the formulation in (12) assumes that the block always starts with noise-only components, and then the signal paths start to arrive. However, in practice, there can be cases where the first step chooses a block consisting of all noise components. By combining a large number of frames; that is, by choosing a large $N_1$ in (7), the probability of this event can be reduced considerably. However, very large $N_1$ also increases the estimation time; hence, there is a trade-off between the estimation error and the estimation time. In order to prevent erroneous TOA estimation when a noise-only block is chosen, a one-sided test can be applied using the known distribution of the noise outputs. Since the noise outputs have a Gaussian distribution, the test reduces to comparing the average energy of the outputs after the estimated change instant to a threshold. In other words, if

$\frac{1}{n_f - \hat{k} + 1} \sum_{i=\hat{k}}^{n_f} z_i^2 < \delta_1$, the block is considered as a noise-only block and the two-step algorithm is run again.

Another improvement of the algorithm can be obtained by checking if the block consists of all signal paths; that is, the TOA is prior to the current block. Again, by following a one-sided test approach, we can check the average energy of the correlation outputs before the estimated TOA against a threshold and detect an all-signal block if the threshold is exceeded. However, for very small values of the TOA estimate $\hat{k}$, there can be a significant probability that the first signal path arrives before the current observation region since the distribution of the correlation output after the first path includes both the noise distribution and the signal plus noise distribution with some probabilities as shown in (13). Hence, the test may not work although the block is an all-signal block. Therefore, some additional correlation outputs before $\hat{k}$ can be employed as well when calculating the average power before the TOA estimate. In other words, if $\frac{1}{\hat{k} - n_s + M_3} \sum_{i=n_s - M_3}^{\hat{k}-1} z_i^2 > \delta_2$, the block is considered as an all-signal block, where $M_3 \geq 0$ additional outputs are used depending on $\hat{k}$. When it is determined that the block consists of all signal outputs, the TOA is expected to be in one of the previous blocks. Therefore, the uncertainty region is shifted backwards, and the change detection algorithm is repeated.

## 4. SIMULATION RESULTS

In this section, we perform simulations in order to evaluate the performance of the estimator over realistic IEEE 802.15.4a channel models [17]. We consider residential and office environments with both line-of-sight (LOS) and non-line-of-sight (NLOS) situations.

In our simulation scenario, the signal bandwidth is 7.5GHz and the frame time of the transmitted training sequence 300ns. Hence, we consider an uncertainty region of 2250 chips, and divide the region into $N_b = 50$ blocks. The number of pulses, over which the correlations are taken in the first and second steps is given by $N_1 = 50$ and $N_2 = 25$, respectively. Also $M_1 = 180$ additional chips prior to the uncertainty region determined by the first step are included in the second step.

The estimator is assumed to have 10 parallel correlators for the second step. In a practical setting, the estimator can use the correlators of a RAKE receiver that is already present for the signal demodulation, and 10 is a conservative value in this sense.

In Figure 2, root mean square errors (RMSEs) are plotted for different signal-to-noise ratios (SNRs) for four different channel models. When the SNR is high enough, very accurate TOA estimates can be obtained. Due to the different characteristics of the channels in residential and office environments, the estimates are better in the office environment. Namely, the delay spread is smaller in the channel models for the office environment. Moreover, as expected, the NLOS situations cause increase in the RMSE values.

Each TOA estimation is performed in approximately $0.92$ millisecond. Because we did not employ any additional tests after the TOA estimate, which are described in Sub-section 3.3, and used the same parameters for all the channel models, the estimation time is the same for all the channel realizations. More accurate results can be obtained by employing different parameters in different scenarios.

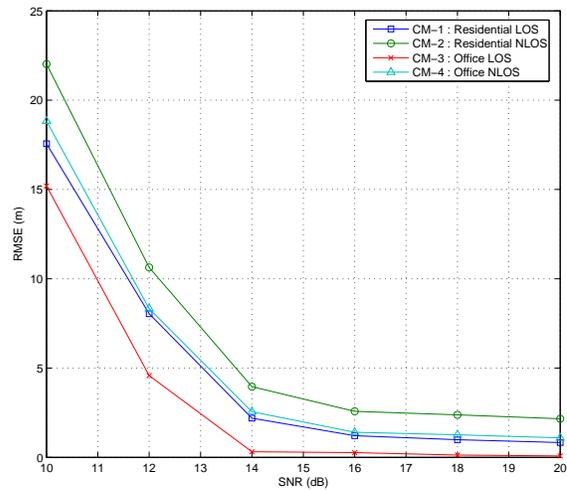

**Fig. 2**. Root mean square error (RMSE) versus signal-to-noise ratio (SNR) for four different IEEE 802.15.4a channel models. The averages over 100 realizations are plotted for each model.

## 5. CONCLUSIONS

In this paper, we have proposed a two-step TOA estimation algorithm, where the first step uses RSS measurements to quickly obtain a rough TOA estimate, and the second step uses a change detection approach to estimate the fine TOA of the signal. The proposed scheme relays on low-rate correlation outputs, but still obtains a considerably accurate TOA estimate in a reasonable time interval, which makes it very practical for UWB systems.

The future work includes optimization of the estimator parameters, such as $N_b$, by considering the statistics of the channel parameters defined by the IEEE 802.15.4a channel committee.